 \documentstyle[NATO,numreferences]{crckapb}

\begin{opening}
\title{Warm Dark Matter}
\subtitle{Clues to Primordial Phase Density from the Structure of Galaxy Halos}  
\author{Craig J. Hogan}
\institute{Astronomy and Physics Departments \\
 University of Washington\\
 Seattle, Washington 98195, USA}
\end{opening}
 \runningtitle{WARM DARK MATTER}
\begin{document}
\begin{abstract}
The Cold Dark Matter paradigm successfully explains many phenomena on scales
larger than galaxies, but seems to predict galaxy halos which are more 
centrally concentrated   and have a lumpier substructure than observed.
Endowing cosmic dark matter with  a small primordial velocity dispersion
preserves  the successful  predictions of the Cold Dark Matter scenario on 
large scales and improves the agreement with halo structure.
A ``phase density''  $Q$, proportional to the inverse entropy for 
nonrelativistic matter, is estimated for relativistically decoupled thermal 
or degenerate relic particles  of mass $m_X$, with a numerical factor  
depending on the particle type but no cosmological parameters. Since $Q$ 
cannot increase for dissipationless, collisionless matter, at a given 
velocity dispersion there is a maximum space density; this
``phase packing'' constraint eliminates the singular density predicted by CDM.
The core radius and  halo circular velocity scale analogously to
degenerate dwarf stars.  Particle velocities also filter primordial 
perturbations on a scale depending on $Q$ and on details of particle
distributions. Particle candidates for warm matter are briefly
discussed; for  warm thermal relics   to have the observed mass density
 requires decoupling prior to the QCD epoch  and therefore a
superweak interaction with thermal Standard Model particles.
\end{abstract}
\newpage
\section{How Cool is Cold Dark Matter?}
The successful match of predictions for large scale structure and
microwave anisotropy vindicates many assumptions of standard cosmology,
in particular the hypothesis that the dark matter is composed of
primordial particles which are cold and collisionless. 
At the same time,  
   the CDM
paradigm finds difficulty explaining   the small-scale structure within galaxy
haloes:     CDM appears to predict excessive relic substructure\cite{ghigna,klypin}
in the form of many dwarf galaxies which 
are not seen and may disrupt disks, and also
predicts\cite{dubinski,nfw,moore98,moore99a,moore99b} a universal, monotonic   increase
of  density towards the center of all halos  which is not seen  in close studies of 
dark-matter-dominated galaxies.  The latter problem   seems to arise
as a generic result of low-entropy material sinking during 
halo formation, quite independently of initial conditions;
the former effect seems to be a generic result of hierarchical clustering predicted
by CDM power spectra, which produce fluctuations on small scales
that collapse early and survive as substructure. 
Although these problems are still controversial from both
a theoretical and observational point of view\cite{frank},
it is not easy to dismiss these effects  by various
complicated baryonic devices.

 It is possible
 that the problems with halo structure are 
  giving specific quantitative clues about new properties of the dark matter particles.
The  existence of dwarf cores and smooth substructure  are just
what one would expect if the dark matter is not absolutely cold but has
 a small nonzero primordial velocity dispersion.
 Such a model produces two
separate effects:  a  phase packing  or Liouville limit  which produces halo cores, and
a filter in the primordial power spectrum which limits small-scale substructure.
The estimated dispersion required for the two effects does not quite agree
but is close enough to motivate a closer look. This discussion is   meant
to motivate more detailed comparison of models and theory  aimed at
  using halo properties to test the hypothesis  of primordial
velocity dispersion and ultimately to measure particle properties.

The physics of the filtering by  freely streaming particles closely parallels that of
massive neutrinos\cite{gerstein,cowsik,bond,bardeen,dodelson},  the standard form of
 ``hot'' dark matter. 
A thermal particle which is more weakly interacting
and therefore separates out of equilibrium earlier than neutrinos,
when there are more particle degrees of freedom, has a lower temperature than neutrinos
and therefore both a lower density and  lower velocity dispersion 
for a given mass. Such ``thermal relics'' constitute one class of warm dark matter
candidate; there are other possibilities, including degenerate 
particles\cite{fuller,shi} and products of decaying particles. In many respects
their astrophysical effects are very similar to the thermal case since
 up to numerical factors the damping scale and phase packing limit are
both fixed by the same quantity, the classical ``phase density'' of the particles.  
For particles which separate out  when relativistic, the phase density 
depends only on the particle properties and not on any cosmological 
parameters. It is similar (though not identical) for bosons and fermions,
and in thermal and degenerate limits.

The phase-packing constraint is  also familiar
from the context of massive neutrinos. Tremaine and Gunn\cite{tremaine}
showed that the phase density of dark matter in giant galaxies implies a 
large neutrino mass, therefore too large a mean cosmic density.  
However, their argument can be turned around to explain the 
lack of a cusp at the center of dwarf halos.  An upper limit to the 
central density of an isothermal sphere can be derived for a given 
phase density; a very rough comparison with dwarf dynamics suggests a
 limit corresponding to that of  a thermal relic with a mass of about 200  
eV. This is lighter (that is, a larger dispersion) than the $\approx$1 keV
currently guessed at from
the filtering effect, but not by so much that the idea should be abandoned;
the simulations and comparison with data may yet reconcile the two effects and
could reveal  correlations  of core radius and velocity dispersion
predicted by the
phase-packing hypothesis.
 Comparison of the two effects may also reveal  new dark matter physics:
for example,  if the particles scatter off each
other by  self-interactions (which may even  be negligible today), free streaming
is suppressed and the filtering  occurs on a somewhat smaller scale.

Warm
dark matter has most often  been invoked as a solution to fixing  
apparent (and no longer problematic\cite{peacock}) difficulties
with predictions of the CDM power spectrum for matching galaxy clustering data.
A  filtered
spectrum may however  solve  several other classic problems of CDM  on smaller
scales, in
galaxy formation itself.  Baryons tend to cool and collapse early into lumps smaller and
denser than observed galaxies\cite{whiterees}. 
 Although this may be prevented by stellar-feedback 
effects,   recent studies\cite{navarrosteinmetz} suggest  that
CDM has fundamental difficulties explaining the basic properties
of galaxies such as the Tully-Fisher relation; dynamical
loss of angular momentum results in halos which are   too concentrated.
Some of these problems may   be solved\cite{sommerlarsen} in warm dark matter
models which suppress the early collapse of subgalactic structures. 
Modeling disk formation includes baryonic evolution so  requires
understanding the ionization history of gas; an important constraint comes
from the observed structures of
the Lyman-$\alpha$ forest\cite{croft}.  Simulations\cite{sommerlarsen}
suggest that the optimal filtering scale corresponds to a thermal particle
mass of about 1 keV.

For these relatively massive thermal relics 
to have the right mean density today, the particle must have
  separated out at least
as early as
 the QCD era, when the number of degrees of freedom was significantly 
larger than at classical weak decoupling. Its 
interactions with normal Standard Model particles must be ``weaker than weak,''
ruling out not only neutrinos but many other particle candidates.
The leading CDM particle
candidates, such as   WIMPs and axions,   form in standard scenarios with
extremely high phase densities, although more elaborate
mechanisms are possible to endow these particles with the required velocities.  It is
therefore of considerable interest from a particle physics point of view to 
find evidence for the existence of finite primordial phase density from galaxy halo
dynamics. Neither the theory or the observational side allow a definitive case to be made
as yet, but the evidence is certainly suggestive. In principle, dynamics can provide
detailed clues to the dark matter mass and interactions. 
 Here I give a few examples of simple calculations which reveal the connections
between the particle properties and the halo properties. 

\section{Phase Density of Relativistically-Decoupled Relics}
Consider particles of mass $m$ originating in   equilibrium and
decoupling at
a temperature $T_D>>m $ or chemical potential $\mu>>m $. The original
distribution function is\cite{landau}
\begin{equation}
f(\vec p)=(e^{(E-\mu)/T_D}\pm 1)^{-1}\approx (e^{(p-\mu)/T_D}\pm 1)^{-1}
\end{equation}
with $E^2=p^2+m^2$ and $\pm$ applies to fermions  and bosons
respectively.  The density and pressure of the particles are\cite{kolb}
\begin{equation}
n ={g\over (2\pi)^3}\int fd^3p
\end{equation}
\begin{equation}
P ={g\over (2\pi)^3}\int {p^2\over 3E} fd^3p
\end{equation}
where $g$ is the number of spin degrees of freedom. Unless stated otherwise,
we adopt units with $\hbar=c=1$.

With adiabatic expansion this distribution
is preserved with  momenta of
particles vary as $p\propto R^{-1}$, so the density
and pressure can be calculated at any subsequent time\cite{peebles}.
For thermal relics $\mu=0$, we can  
derive the density and pressure in the   limit
when the particles have cooled to be nonrelativistic:
\begin{equation}
n={g T_0^3\over (2\pi)^3}\int  {d^3p \over e^p\pm 1}
\end{equation}
\begin{equation}
P={g T_0^5\over (2\pi)^3 3m}\int   {p^2d^3p \over e^p\pm 1}
\end{equation}
where the  pseudo-temperature $T_0=T_D(R_D/R_0)$ records the expansion of
any fluid element. 

It is useful to define a ``phase density'' $Q\equiv
 \rho/\langle v^2\rangle^{3/2}$ proportional
to the inverse specific entropy for nonrelativistic matter. 
This quantity is preserved under adiabatic expansion but for
nondissipative particles cannot increase. 
Combining the above expressions for density and pressure
and using $\langle v^2\rangle=3P/nm$, we find
\begin{equation}
Q_X =q_X  g_X m_X^4.
\end{equation}
 The coefficient for the thermal case is 
\begin{equation}
q_T={4\pi  \over (2\pi)^3}
{ [\int dp (p^2/e^p\pm 1)]^{5/2}    \over
[\int dp (p^4/e^p\pm 1)]^{3/2}     }=0.0019625,
\end{equation}
where the last equality holds for thermal fermions.
An analogous calculation for the degenerate fermion case 
($T=0, \mu_D>>m_X$) yields the same expression for
$Q $ but with a different coefficient,
\begin{equation}
q_d={4\pi  \over (2\pi)^3}
{ [\int_0^1p^2 dp]^{5/2}    \over
[\int_0^1p^4 dp ]^{3/2}     }=0.036335.
\end{equation}

The phase density depends on the particle properties but not
at all on the cosmology; the decoupling temperature, the current
temperature and density  do not matter. Up to numerical factors 
 (which depend on thermal or degenerate, boson or fermion
cases), the phase density for relativistically 
decoupled or degenerate matter is set just  by the particle mass $m_X$.

\section{Phase Packing and the Core Radius of an Isothermal Halo}

For a given velocity dispersion at any point in space, the primordial phase density
of particles imposes an upper limit on their density $\rho$.
Thus dark matter halos do not form the singular central cusps
predicted by Cold Dark Matter but instead form cores with constant 
  density   at small radius.  A lower limit to the size of the core
 can be estimated
if we assume that the matter in the central parts of the halo 
lies close to the primordial adiabat defined by $Q$. This
will be  good model for cores which form quietly without too much
dynamical heating. This seems to be the case in typical CDM halos,
indicated by the cusp prediction; it could
be that warm matter typically experiences more additional
dynamical  heating than cold matter,
in which case the   core could be larger.

The conventional definition of core size in an isothermal
sphere\cite{binney} is the 
``King radius''
\begin{equation}
r_0=\sqrt{9\sigma^2/ 4\pi G \rho_0}
\end{equation}  
where $\sigma$ denotes  the one-dimensional velocity dispersion
and $\rho$ denotes the central density.  Making the adiabatic
assumption, $\rho_0= Q (3\sigma^2)^{3/2}$, we find
\begin{equation}
r_0=
\sqrt{9 \sqrt{2}/ 4\pi 3^{3/2}}(QG v_{c\infty})^{-1/2}=
0.44 (QG v_{c\infty})^{-1/2}
\end{equation} 
(Note that aside from numerical factors this is the same  as
a degenerate dwarf star; the galaxy
core is bigger than a Chandrasekhar dwarf of the same specific binding energy by a
factor
$(m_{proton}/m_X)^2$.)
For comparison with observations we have expressed 
the core radius as a function of the asymptotic circular velocity
$v_{c\infty}=\sqrt{2}\sigma$
of the halo's flat rotation curve.

For the thermal and degenerate phase densities
derived above, 
\begin{equation}
r_{0,thermal}= 5.5 {\rm kpc} (m_X/100{\rm eV})^{-2}
(v_{c\infty}/30{\rm km s^{-1}})^{-1/2}
\end{equation}
\begin{equation}
r_{0,degenerate}= 1.3 {\rm kpc} (m_X/100{\rm eV})^{-2}
(v_{c\infty}/30{\rm km s^{-1}})^{-1/2},
\end{equation}
where we have set $g=2$.
The circular velocity in the central
core displays the harmonic behavior $v_c\propto r$; it reaches half of
its asymptotic  value at a radius $r_{1/2}\approx 0.4 r_0$.

The best venue for studying the effect is in the small, dark-matter-dominated
disk galaxies where the cusp problem of CDM seems most clearly defined. Material on
circular orbits gives a direct measure of the enclosed mass and therefore of the density
profile.  A rough guess from current
 observations of inner rotation curves of a few dwarf
spiral galaxies  suggests a halo core corresponding to a thermal particle mass  of
about 200 eV\cite{dalcanton}; for a larger mass,
additional (nonprimordial) dynamical heating is required. This is also
consistent with what is known about   dark matter
in dwarf elliptical galaxies from studies of
stellar velocities\cite{aaronson,olszewski,mateo}. 

  The relationship of core radius with halo velocity
dispersion is a simple prediction of the primordial phase density
explanation of cores which will probably generalize in some form to 
a cosmic population of halos. In particular if
phase packing is the explanation of dwarf galaxy cores, the dark matter cores of giant
galaxies and galaxy clusters  are predicted to be much smaller than for dwarfs, 
   unobservably hidden in a central region dominated by
baryons. 

\section{Filtering  of Small-Scale Fluctuations}
The transfer function of Warm Dark Matter is almost
the same as Cold Dark Matter on large scales, but is
filtered by free-streaming on small scales. The 
characteristic wavenumber for filtering at any time is given
by $k_X\approx H/\langle v^2\rangle^{1/2}$, with a filter shape
depending on the detailed form  of the distribution function.
In the current application,
we are concerned with $H$ during the radiation-dominated era
($z\ge 10^4$), so that $H^2=8\pi G \rho_{rel}/3\propto (1+z)^2 $, where 
$\rho_{rel }$ includes all relativistic degrees of freedom.
For constant $Q$, $\langle v^2\rangle^{1/2}= (\rho_X/Q)^{1/3}\propto
(1+z)$   as long as the $X$ particle
are nonrelativistic. The  maximum  
comoving filtering scale\cite{kolb} is thus approximately independent
of redshift and is given simply by
\begin{equation}
k_X\approx H_0 \Omega_{rel}^{1/2} v_{X0}^{-1}
\end{equation}
where 
$\Omega_{rel}= 4.3\times 10^{-5}h^{-2}$ and  
 $v_{X0}=(Q/\bar\rho_{X0})^{-1/3}$ is the rms velocity of
the particles at their present mean cosmic density
$\bar\rho_{X0}$: 
\begin{equation}
k_X\approx 0.65 {\rm Mpc^{-1}} (v_{X0}/1 {\rm km\ s^{-1}})^{-1},
\end{equation}
with no dependence on $H_0$.
For the thermal and degenerate cases, in terms of particle
mass we have
\begin{equation}
v_{X0,thermal}=0.93 {\rm km\ s^{-1}} h_{70}^{2/3} (m_X/ 100 {\rm
eV})^{-4/3}(\Omega_X/0.3)^{1/3}g^{-1/3}
\end{equation}
\begin{equation}
v_{X0,degenerate}=0.35 {\rm km\ s^{-1}} h_{70}^{2/3} (m_X/ 100 {\rm
eV})^{-4/3}(\Omega_X/0.3)^{1/3}g^{-1/3}
\end{equation}

In the case of free-streaming, relativistically-decoupled
thermal particles, the transfer function has been computed
precisely\cite{bardeen,sommerlarsen}; the characteristic
wavenumber where the square of the transfer function falls to half 
the CDM value is $k_{1/2,stream}= k_X/5.5$. The mass implied for this
kind of candidate to preserve the success of CDM on galaxy scales
and above is about\cite{sommerlarsen} 1 keV; if it is much smaller (in particular, 
as small as the 200 eV we require for phase packing alone
to help the core problem), filtering occurs on too large
a scale. 
(A filtering scale  of roughly $k\approx 3 h_{70} {\rm Mpc}^{-1}$ preserves the 
successes  of CDM on large  scales and
  helps to solve  the CDM predictions of excess
dwarf galaxies, excessive substructure in halos, insufficient angular momentum, and
excessive baryon concentration.) This problem
might be fixed in other models with a different relationship of
$k_{1/2}$ and $k_X$. For example, if the particles are self-interacting 
the free streaming is suppressed and the relevant scale is
the standard Jeans scale for acoustic oscillations,
$k_J=\sqrt{3}H/c_S=\sqrt{27/5}k_X$, which is significantly shorter
than  $k_{1/2,stream}$ at a fixed phase density.  Alternatively it is
possible that warm models might be more effective at producing smooth cores
than we have guessed from the minimal phase-packing  constraint
above; an evaluation of this possibility requires simulations which
include not just a filtered spectrum but a reasonably complete
sampling of a warm distribution function in the particle velocities\cite{wadsley}. 

\section{Density of Thermal Relics}

A simple  candidate for warm dark matter is a  standard
thermal relic--- a particle that decouples from the thermal background very
early, while it is still relativistic. 
In this case the mean density of the particles can be estimated\cite{kolb}
from   the
number of  particle degrees of freedom at the epoch $T_D$ of decoupling,
$g_{*D}$:
\begin{equation}
\Omega_X =7.83 h^{-2} [g_{eff}/g_{*D}] (m/{\rm 100 eV})
=0.24 h_{70}^{-2} (m_X/{\rm 100 eV}) (g_{eff}/1.5)(g_{*D}/100)^{-1} 
\end{equation} 
where $g_{eff}$  is
the number of effective photon degrees of freedom of the particle  
($=1.5$ for a two-component fermion).
For standard neutrinos which decouple at around 1MeV, $g_{*D}=10.75$.

An acceptable mass density for a warm relic with $m_X\ge$ 200 eV clearly requires a much
larger 
$g_{*D}$ than the standard value for neutrino decoupling. Above about 200 MeV,   the 
 activation of the extra gluon and quark degrees of freedom
(24 and 15.75 respectively including $uds$ quarks)
give  $g_{*D}\approx 50$;  activation of heavier modes of the Standard
Model above $\approx 200$GeV produces 
$g_{*d}\approx 100$ which give  a better match for $\Omega_X\le 0.5$, as suggested
by current evidence.  Masses of the order of 1 keV can be accomodated
by adding extra, supersymmetric degrees of freedom.
Alternatively a  degenerate particle can be introduced via mixing of a sterile
neutrino, combined with a primordial chemical potential adjusted to give the right
density\cite{shi}.
  Either way, the particle must interact
with Standard Model particles much more weakly than normal weak interactions,
which decouple at $\approx 1$ MeV. 

Note that all warm dark matter particles have low 
 densities compared with photons and other species at 1 MeV
 so they do  not strongly affect 
nucleosynthesis. However, their effect is not entirely negligible.
They add   the
equivalent of $(T_X/T_\nu)^3= 10.75/g_{*d}\approx$
 0.1 to 0.2 of an effective extra neutrino species,
which leads to a small increase in the predicted primordial helium abundance  for a
given 
$\eta$. Because the phase density fixes
the mean density at which the particles become relativistic, this
is a generic feature for any warm  particle.  This effect might
become detectable with increasingly precise measurements of cosmic abundances.

\acknowledgements
I am grateful for useful discussions  of these issues with
F. van den Bosch, J. Dalcanton, A. Dolgov,
G. Fuller, B. Moore, J. Navarro,  T. Quinn, J. Stadel,
J. Wadsley, and S. White, and for the 
hospitality  of the Isaac Newton Institute for Mathematical
Sciences  and the Ettore Majorana Centre for Scientific
Culture.
This work was supported at the University of Washington
by NSF and NASA, and at the Max-Planck-Institute f\"ur
Astrophysik by a Humboldt Research Award.

\end{document}